\documentclass[a4paper]{IEEEtran} 

\usepackage{csvsimple}

\usepackage{array}

\usepackage{mdwmath}
\usepackage{mdwtab}

\usepackage{flushend}           
\usepackage{fixltx2e}

\usepackage{url}
\usepackage{times}
\usepackage[pdftex]{graphicx}	

\usepackage[numbers,sort&compress]{natbib}

\usepackage{versions}

\usepackage[capitalize]{cleveref}

\usepackage[caption=false,font=footnotesize]{subfig}

\usepackage{paralist}           

\usepackage{algorithmic}       
\usepackage{listings}           
\usepackage{moreverb}           

\usepackage{fancyvrb}           

\usepackage{eurosym}
\usepackage{units}              

\usepackage{multicol}           
\usepackage{longtable}          
\usepackage{booktabs}
\usepackage{ctable}             



\usepackage{rotating}           

\usepackage{multirow}

\newcommand{\PreserveBackslash}[1]{\let\temp=\\#1\let\\=\temp}

\usepackage[normalem]{ulem}     

\usepackage{xspace}
\usepackage{xcolor}
\definecolor{pink}{rgb}{1,.4,.4}
\definecolor{red}{rgb}{1, .2, .1}

\usepackage{fp}                 

\newcommand{\GSD}{GSD\xspace}
\newcommand{\GSDlong}{Global Software Development\xspace}

\newcommand{\nTotal}{64}

\newcommand{\nUnneeded}{7}

\FPsub{\nRelevantf}{\nTotal}{\nUnneeded}
\FPround{\nRelevant}{\nRelevantf}{0}

\begin{document}
\title{An Empirical Study on \emph{Leanness} and \emph{Flexibility} in Distributed Software Development}
\author{
  \IEEEauthorblockN{Mohammad Abdur Razzak\\}
  \IEEEauthorblockA{
    Lero, the Irish Software Research Centre, 
    University of Limerick, Ireland\\
    \path|{abdur.razzak}@lero.ie|\\
  }
}

\maketitle
\thispagestyle{empty}

\begin{abstract}

Nowadays, many individuals and teams involved on projects are already using agile development techniques as part of their daily work. However, we have much less experience in how to scale and manage agile practices in distributed software development. Distributed and global development-- that requiring attention to many technical, organizational, and cultural issues as the teams interact to cooperatively delivery the solution. Alongside, very large team sizes, teams of teams, and more complex management structures forcing additional attention to coordination and management. At this level, there is an increasing need to standardize best practices to avoid reinvention and miscommunication across artifacts and processes. Complexity issues in enterprise software delivery can have significant impact on the adoption of agile approaches. As a consequence, agile strategies will typically need to be evaluated, tailored, and perhaps combined with traditional approaches to suit the particular context. The characteristics of software products and software development processes open up new possibilities that are different from those offered in other domains to achieve leanness and flexibility. Whilst Lean principles are universal, a further understanding of the techniques required to apply such principles from a software development angle. Thus, the aim of this research is to identify, how leanness facilitate flexibility in distributed software development to speed-up development process.


\end{abstract}

\section{Background}\label{sec:background}
	

\subsection{Global Software Development}
Improved communication technologies, access to global talent, cheaper labour, proximity to new markets and legal requirements have all contributed to the growth in \GSDlong (\GSD) \cite{vizcaino2016validated}. \GSD is software work undertaken in different geographical locations, across national boundaries in a coordinated fashion through synchronous and asynchronous interaction \cite{herbsleb2001global}. As a result, a growing numbers of software companies started to implement \GSD to reduce time-to-market, increase operational efficiency, improve quality, and many more. Over the years several recommendations have been published in support of this complex development paradigm \cite{beecham2013who}. But, industrial experience shows that, \GSD is reputed to suffer from communication breakdowns, low morale and delays due to teams being geographically, culturally and temporally separated \cite{noll2010global, beecham2014motivating, beecham2015motivates, conchuir2009global}.



\subsection{Agile Methods}

Traditionally \GSD has followed a plan driven, structured, waterfall approach, where tasks are allocated according to where they appear in the software lifecycle\cite{estler2014agile}. It was considered that agile methods envisaged for \emph{small projects} and \emph{co-located teams} \cite{kahkonen2004agile, abrahamsson2009lots} would be a poor fit for \GSD because both Agile and distributed development approaches differ significantly \cite{ramesh2006can}. Agile methods tend to rely on informal processes to facilitate coordination whereas distributed software development relies on formal mechanisms. There is a growing trend for companies to adopt agile methods as reported in a tertiary study of \GSD \cite{hanssen2011signs}. Adopting Agile practices such as short iterations, frequent builds, and continuous delivery all pose challenges to configuration management and version management \cite{paasivaara2006could}. But, practices such as \emph{Short iterations} increase transparency of Work-in-Progress (WIP) and provde a big picture project progress to stakeholders \cite{paasivaara2004using}. However, setting up an Agile team is usually motivated by benefits such as increased productivity, innovation, and employee satisfaction \cite{smite2010fundamentals} but introducing an Agile method can change the culture (command and control model) in a company; so to implement the Agile practices in global software environment developers need to have more autonomy as well as decision-making power \cite{fowler2006using}.


\subsection{Lean in Software Development}
Lean was born as part of the industrial renaissance in Japanese manufacturing after the \emph{Second World War} in the 1940s but the team ``Lean" was first applied publicly to a production management process and then to product development at MIT during the mid-1980s; a detailed description of the story of Lean can be found in the book \emph{``The machine that changed the world"} \cite{womack1990machine}. In general, Lean is a manufacturing \& production practice that considers the expenditure of resources for any goal other than the creation of value for the end customers to be wasteful, and thus a target for elimination. In 2003, Poppendieck et al. \cite{poppendieck2003lean} illustrated how many of the lean principles and practices can be used in Software Engineering context. Lean Software Development (LSD) shares principles with Agile especially people management and leadership, quality and technical excellence, and frequent and fast delivery of value to the customers \cite{petersen2011measuring}.

The core five principles of Lean thinking according to MIT's researchers are \cite{womack1996lean}: \emph{Value}, \emph{Value Stream}, \emph{Flow}, \emph{Pull}, and \emph{Perfection}. But, it is challenging to adopt those principles in software development due to domain variability i.e; manufacture \cite{munch2012software}. The concept of value is not straightforward in software development because it is not limited to a single time-bound effort \cite{poppendieck2012lean}. Waste is also a controversial matter as work items in software development are much more intangible. The principles of value stream and flow are also challenged because software development is a process that bases mainly on information. Software development is an knowledge intensive job which relies on creativity, knowledge, and experience. That means, human factor is a dominant factor in software development whilst in a manufacturing environment human presence is mainly required to operate automated machines. The main goal is to implement lean manufacturing principles into a software development model is to reduce the waste in a system and produce a higher value for the final customer. Poppendieck et al. \cite{poppendieck2012lean} mentioned, \emph{``If lean is thought of as a set of principles rather than practices, then applying lean concepts to product development and software engineering makes more sense and can lead to process and quality improvements"}.

According to Poppendieck and Poppendieck \cite{poppendieck2012lean,poppendieck2003lean} interpretations of Lean thinking in Software Development, there are seven principles that guide Lean Software Development as follows:

\begin{itemize}

\item Eliminate waste, understanding first what value is.
\item Build quality in, by testing as soon as possible, automation and refactoring.
\item Create knowledge, through rapid feedback and continuous improvement.
\item  Defer commitment, by maintaining options and making irreversible decisions in the last responsible moment when most information is available. 
\item  Deliver fast, through small batches and limiting WIP.
\item Respect people, the people doing the work.
\item Optimise the whole, by implementing Lean across an entire value stream.
Seven sources of waste in software development: partially done work, extra features, relearning, handoffs, task switching, delays and defects.

\end{itemize}

The interpretations presented above can be considered in practice as the body of knowledge of Lean Software Development.

\subsection{Combining Lean and Agile}

Scaling Agile continues to be a challenge in software development. Lean software development is acquiring an identity of its own as a means to scale Agile. But, Wang and Conboy \cite{wang2011comparing} question whether Agile and Lean are just two different names for the same thing, or whether they are actually different and, therefore, the challenges and issues faced by Agile processes could be addressed by Lean approaches. On the other hand, Petersen concluded that, both paradigms share almost same principles such as managing people, continuous attention to quality and technical excellence. However, the end-to-end focus and flow are unique to Lean. Different literatures also claimed that, empirical studies are need to identify the difference between Agile and Lean or combination of both in software development \cite{wang2011comparing}. In general, the most important goals for Agile and Lean adopters are \cite{rodriguez2012survey}:  

 \begin{itemize}
\item To reduce development cycle times and time-to-market
\item To improve process quality
\item To remove waste and excess activities

\end{itemize}


\section{Motivation}\label{sec:motivation}



Nowadays, many individuals and teams involved on projects are already using agile development techniques as part of their daily work. However, we have much less experience in how to scale and manage agile practices in distributed software development. According to Alan W. Brown~\cite{brown2011case}, the one of the top most complex scaling agile issue is \emph{Distributed and global development-- that requiring attention to many technical, organizational, and cultural issues as the teams interact to cooperatively delivery the solution}. The author also mentioned, very large team sizes, teams of teams, and more complex management structures forcing additional attention to coordination and management. At this level, there is an increasing need to standardize best practices to avoid reinvention and miscommunication across artifacts and processes.

Scaling agile means moving from few agile teams to multiples or even more such as hundreds of agile development teams. Scaling Agile continues to be a challenge is software development because when more teams works together then its required strong coordination among teams as well as on the project \cite{turk2014limitations, pikkarainen2008impact, maples2009enterprise, abrahamsson2009lots}. Scott W. Ambler~\cite{ambler2008agile} pointed out several factors, that needs to consider when scaling Agile such as team size, geographical distribution, entrenched culture, system complexity, legacy systems, regulatory compliance, organizational distribution, governance and enterprise focus. In general, productivity and quality are the two main concern of any organization to adopt the concept of scaling agile.

Complexity issues in enterprise software delivery can have significant impact on the adoption of Agile approaches~\cite{brown2011case}. As a consequence, Agile strategies will typically need to be evaluated, tailored, and perhaps combined with traditional approaches to suit the particular context. The characteristics of software products and software development processes open up new possibilities that are different from those offered in other domains to achieve leanness and flexibility. Whilst Lean principles are universal, a further understanding of the techniques required to apply such principles from a software development perspective. Thus, the aim of this research is to identify, how leanness\footnote{Lean Software Development} facilitate flexibility\footnote{Agile Software Development} in distributed software development to speed-up\footnote{Continuous delivery and time-to-market} development process.


\section{Research Questions}\label{sec:questions}

\begin{enumerate}

\item  Are agile practices useful in distributed software development context? What challenges and open issues arise with their introduction? 
\item How flexibility and leanness can be combined to speed-up distributed software development?
\item What are the variability factors in scaling Lean and Agile in distributed software development? 



\end{enumerate}


\section{Method}\label{sec:method}


\subsection{Research Methods}

\begin{figure}[htb] \centering{
\includegraphics[scale=0.4]{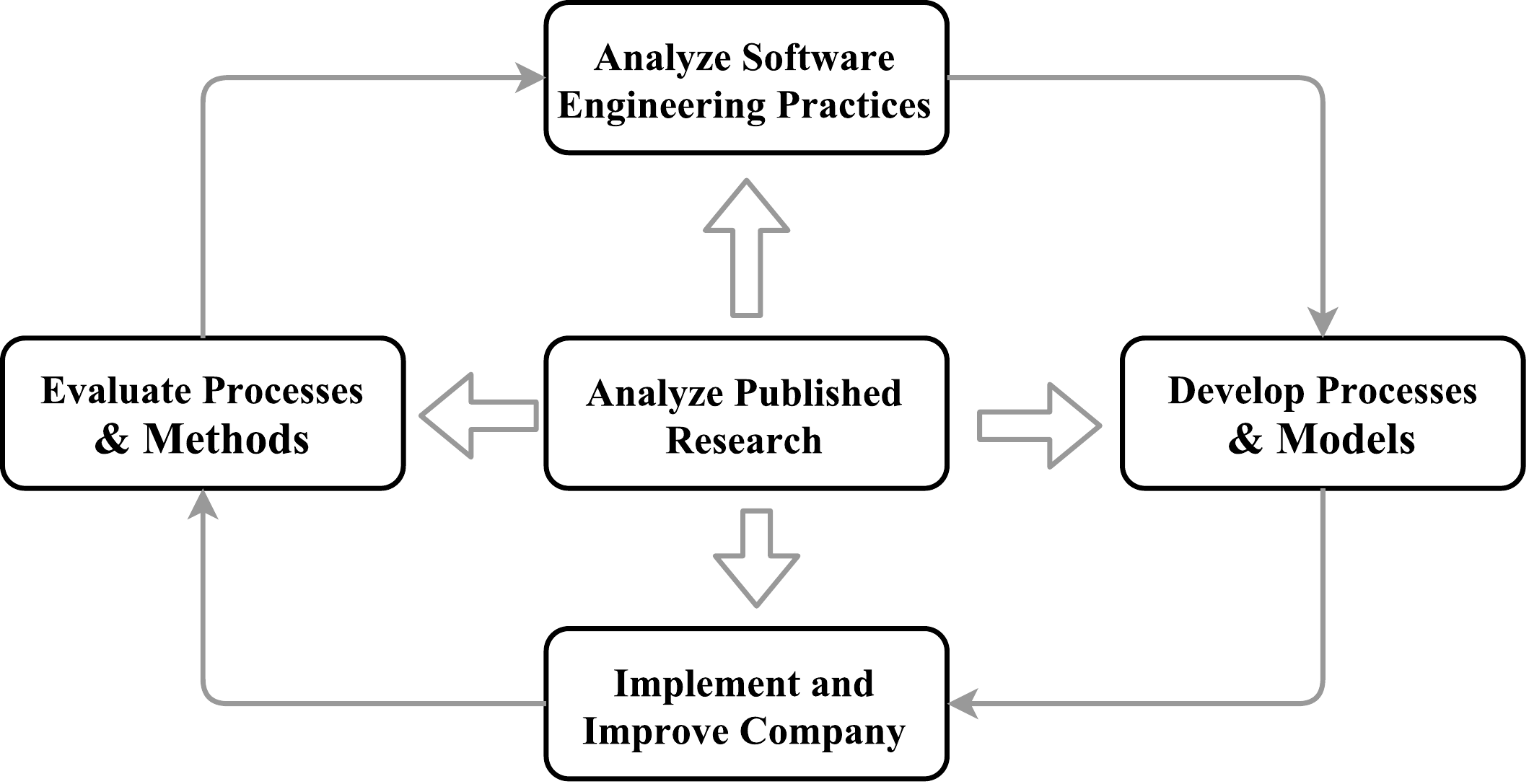}}
\caption{Action Research}
\label{fig:ar}
\end{figure}

This research will undertake a cycle of action research. According to Robson, the purpose of action research is to \emph{``influence or change some aspect of whatever is the focus of the research"}. In action research, he researcher is actively involved in introducing the intervention and making the observations \cite{colin2002real}, in fact the researcher takes an active part in the organization (e.g. by participating in a development team affected by the intervention introduced). As pointed out by Martella et al. \cite{martella1999research} much can be learned by continuously observing the effect of a change after inducing it. However, as the researcher is actively involved in the team work action research is an effort intensive approach from the researcher's point of view. Action research involves close cooperation between practitioners and researchers to bring about change. The action research process can be defined as a number of learning cycles consisting of predefined stages, as presented in Fig. \ref{fig:ar}. 

Within the action research a number of sub-methods will be use, namely interviews and workshops for data collection, and grounded theory as well as statistical analysis. Alongside, we will also gather data through participant observations, informal meetings, informal communications (e-mails) and documents from the organization and specific projects.


This research will comprise multiple iterations over five phases in three stages: 

\begin{itemize}

\item \textbf{Stage 1}
	\begin{itemize}
	\item \emph{Phase 1}: Identify ``Open Issues" of Scaling Lean and Agile in Distributed Software Development

	\item \emph{Phase 2}: Identify the current ``As-is" process in the industrial settings

	\end{itemize}

\end{itemize}

\begin{itemize}

\item \textbf{Stage 2}
	\begin{itemize}
	\item \emph{Phase 3}: We will develop a process implementation ``Roadmap" based on the outcomes, that documents how to transition from the current ``As-is" process, to the desired ``To-be" process.  
	\end{itemize}

\end{itemize}

\begin{itemize}

\item \textbf{Stage 3}
	\begin{itemize}
	\item \emph{Phase 4}: In this stage, we will implement ``To-be" process within the industrial settings and collect the KPI's 
	\item \emph{Phase 5}: In this phase, we will evaluate the implementation and revise Roadmap and ``To-be" models accordingly.
	\end{itemize}

\end{itemize}

\begin{table}[htbp]
\caption{Research Activities}
\centering
\begin{tabular}{ll}
\hline
\textbf{\emph{Research Question}} & {\textbf{\emph{Research Method}}} \\\hline

What makes Lean and Agile development   &  \\ practices successful in GSD? & Literature Survey  \\\hline 

How flexibility and leanness can be combined to  & Action Research \\ speed-up distributed software development? &  \\\hline

What are the variability factors in scaling Lean & Action Research \\ and Agile in distributed software development? &   \\\hline

\end{tabular}

\label{tab:activities}
\end{table}





\textbf{Acknowledgments}
I would like to thanks my supervisor Dr. John Noll, Research Fellow, Lero. This work was supported, in part, by Science Foundation Ireland grants
10/CE/I1855 and 13/RC/2094  to Lero - the Irish Software Research Centre
(\url{www.lero.ie}).

\newcommand{\BIBdecl}{\setlength{\itemsep}{0pt}}

\bibliographystyle{plain}
\bibliography{top}


\end{document}